\documentclass[useAMS]{mn2e}
\usepackage{graphicx}
\usepackage{longtable}

\title[AGB Stars in the Leo I Dwarf Spheroidal Galaxy]{Asymptotic Giant
Branch Stars in the Leo I Dwarf Spheroidal Galaxy} 
\author[Menzies et al.]{John W. Menzies$^{1}$,
Patricia A. Whitelock$^{1,2}$,
Michael W. Feast$^{1,2}$,
and Noriyuki \newauthor Matsunaga$^3$\\
      $^1$ South African Astronomical Observatory, P.O.Box 9, 7935
           Observatory, South Africa.\\
      $^2$ Astronomy Department, University of Cape Town, 7701 Rondebosch,
           South Africa.\\
      $^3$ Institute of Astronomy, University of Tokyo,
           2-21-1 Osawa, Mitaka, Tokyo 181-0015, Japan.\\
 }
\begin{document}
\maketitle

\begin{abstract} 
  Twenty six Asymptotic Giant Branch (AGB) variables are identified in the Local
Group galaxy Leo~I. These include 7 Mira and 5 semi-regular variables for
which periods, amplitudes and mean magnitudes are determined. The large
range of periods for the Miras, $158 <P < 523$ days, suggests an AGB
spanning a significant age range. The youngest must be around 1.6\,Gyr while
the oldest could be 10 Gyr or more. Two of these old Miras are found in the
outer regions of Leo~I (over 490 arcsec from the centre) where stars on the
extended AGB are rare. They could provide an interesting test of third
dredge-up and mass loss in old stars with low metallicity and are worth
further detailed investigation. At least two stars, one a Mira, the other an
irregular variable, are undergoing obscuration events due to dust ejection.

An application of the Mira period-luminosity relation to these stars yields
a distance modulus for Leo~I of $(m-M)_0=21.80\pm0.11$ mag (internal), $\pm
0.12$ (total) (on a scale that puts the LMC at 18.39 mag) in good agreement
with other determinations.
 
\end{abstract}

\section{Introduction}
 This paper is one of a series aimed at finding and characterizing luminous
Asymptotic Giant Branch variables within Local Group galaxies; it
follows similar work on Phoenix and Fornax (Menzies et al. 2008; Whitelock
et al. 2009). Menzies et al.  (2002 hereafter Paper~I) reported the
discovery of several dust-enshrouded AGB stars in Leo~I. Here we discuss
multi-epoch $JHK_S$ photometry of those stars which enables us to
characterise them. The large-amplitude, or Mira, variables are of particular
interest, first, because they tell us about the intermediate age population
of which they are the most luminous representatives, secondly because they
provide an independent distance calibration, via the period-luminosity (PL)
relation (Whitelock et al. 2008), and thirdly because they will be major
contributors to the processed material currently entering the interstellar
medium of Leo~I and are therefore an important source of enrichment.

Leo~I is one of the most distant of the Local Group Galaxies associated with
the Milky Way. It has a complex star formation history (Gallart et al. 1999;
Hernandez et al. 2000; Dolphin 2002) with evidence for star formation over
much of its lifetime. The most recent episode occurred about 1\,Gyr ago and
may have been the result of Leo~I's interaction with the Milky Way (Mateo,
Olszewski \& Walker 2008). Most published work indicates little evidence for
population gradients (e.g. Koch et al. 2007). However, Mateo et al. (2008)
find that the AGB stars are almost exclusively located within a radius of
400 arcsec whereas giant branch (GB) stars have a more extended
distribution. The analysis of infrared photometry by Held et al. (2010) also
indicates a strong radial gradient in the intermediate-age populations and
supports the findings by Mateo et al.

\begin{center}
\onecolumn
\begin{longtable}{ccccccccl}
\caption[]{Probable AGB Stars in Leo~I (excluding variable stars).}
\label{tab_AGB} \\
\hline
\multicolumn{1}{c}{$\alpha$} & $\delta$ & L\# & $J$ & $H$ & $K_S$ & $J-K_S$ & Memb & other name\\
\multicolumn{2}{c}{(equinox 2000.0)}& & \multicolumn{4}{c}{-------------- (mag) -----------------}& & \\
\hline
\endfirsthead

\hline
\multicolumn{1}{c}{$\alpha$} & $\delta$ & L\# & $J$ & $H$ & $K_S$ & $J-K_S$ & Memb & other name\\
\multicolumn{2}{c}{(equinox 2000.0)}& & \multicolumn{4}{c}{-------------- (mag) -----------------}& & \\
\hline
\endhead

  \multicolumn{9}{l}{{Continued on Next Page\ldots}} \\
\endfoot
\multicolumn{9}{l}{Notes: the ``other names" starting with MOW come from Mateo et al. (2008), and with ALW}\\
\multicolumn{9}{l}{from Azzopardi et al. (1986). An ``m" in the Memb column indicates a radial velocity member}\\ 
\multicolumn{9}{l}{according to Mateo et al. The carbon star, ALW7, is incorrectly identified by Demers \& Battinelli } \\ 
\multicolumn{9}{l}{(2002) with their star 10. }\\  

\hline
\endlastfoot

10:08:02.7 & 12:17:56 & 6010 & 16.00 & 15.29 & 15.04 & 0.96 & &  \\
10:08:07.9 & 12:15:18 & 6014 & 16.36 & 15.52 & 15.35 & 1.01 & m & MOW253 \\
10:08:10.7 & 12:17:24 & 6020 & 16.66 & 15.87 & 15.74 & 0.92 & m & MOW167 \\
10:08:10.9 & 12:12:40 & 7019 & 16.70 & 15.97 & 15.89 & 0.81 & &  \\
10:08:14.5 & 12:18:02 & 1024 & 16.07 & 15.27 & 15.07 & 1.00 & m & MOW104 \\
10:08:14.5 & 12:18:02 & 6011 & 16.10 & 15.28 & 15.08 & 1.02 & &  \\
10:08:14.9 & 12:14:04 & 7012 & 15.76 & 15.05 & 14.89 & 0.87 & &  \\
10:08:15.7 & 12:18:14 & 1069 & 16.70 & 15.98 & 15.80 & 0.90 & &  \\
10:08:15.7 & 12:18:14 & 6022 & 16.71 & 16.00 & 15.80 & 0.90 & &  \\
10:08:15.8 & 12:20:32 & 1098 & 16.40 & 15.66 & 15.48 & 0.93 & m & MOW124 \\
10:08:15.8 & 12:20:32 & 6026 & 16.42 & 15.65 & 15.47 & 0.95 & &  \\
10:08:16.6 & 12:20:07 & 1093 & 16.41 & 15.69 & 15.45 & 0.97 & m & MOW101,ALW19 \\
10:08:17.0 & 12:18:15 & 1070 & 16.67 & 15.92 & 15.75 & 0.92 & m & MOW78 \\
10:08:17.1 & 12:18:52 & 1078 & 16.56 & 15.80 & 15.64 & 0.92 & m & MOW77 \\
10:08:17.5 & 12:16:15 & 1054 & 16.99 & 16.20 & 16.02 & 0.97 & m & MOW110 \\
10:08:18.0 & 12:20:09 & 1038 & 16.13 & 15.34 & 14.98 & 1.15 & &  \\
10:08:19.4 & 12:17:07 & 1056 & 16.44 & 15.70 & 15.40 & 1.04 & m & MOW67,ALW1 \\
10:08:20.2 & 12:17:58 & 1010 & 15.28 & 14.50 & 14.32 & 0.96 & &  \\
10:08:20.4 & 12:17:45 & 1061 & 16.97 & 16.20 & 16.02 & 0.95 & m & MOW43 \\
10:08:21.3 & 12:14:43 & 1016 & 16.00 & 15.21 & 14.96 & 1.05 & m & MOW161 \\
10:08:21.4 & 12:13:49 & 8023 & 16.38 & 15.61 & 15.37 & 1.01 & m & MOW193 \\
10:08:22.0 & 12:18:49 & 1030 & 16.25 & 15.48 & 15.32 & 0.94 & &  \\
10:08:22.0 & 12:19:37 & 1088 & 16.56 & 15.80 & 15.61 & 0.95 & m & MOW38 \\
10:08:23.2 & 12:18:38 & 1029 & 16.46 & 15.69 & 15.50 & 0.96 & m & MOW12 \\
10:08:23.9 & 12:17:08 & 1057 & 16.41 & 15.63 & 15.48 & 0.94 & m & MOW36 \\
10:08:24.5 & 12:20:40 & 1039 & 16.20 & 15.45 & 15.23 & 0.96 & m & MOW65 \\
10:08:24.7 & 12:23:23 & 4012 & 16.35 & 15.61 & 15.25 & 1.10 & &  \\
10:08:24.8 & 12:19:27 & 1033 & 15.71 & 14.92 & 14.58 & 1.14 & & ALW7 \\
10:08:27.6 & 12:17:32 & 1059 & 16.41 & 15.69 & 15.40 & 1.01 & m & MOW13,ALW10 \\
10:08:27.8 & 12:17:55 & 1062 & 16.59 & 15.81 & 15.62 & 0.97 & m & MOW6 \\
10:08:28.2 & 12:23:30 & 4006 & 16.34 & 15.51 & 15.26 & 1.08 & &  \\
10:08:28.3 & 12:20:29 & 1097 & 16.98 & 16.25 & 16.05 & 0.93 & &  \\
10:08:28.4 & 12:18:49 & 1075 & 16.76 & 16.05 & 15.91 & 0.84 & m & MOW4 \\
10:08:28.7 & 12:16:25 & 1018 & 16.47 & 15.70 & 15.50 & 0.97 & m & MOW57 \\
10:08:29.2 & 12:18:32 & 1026 & 16.14 & 15.39 & 15.18 & 0.96 & m & MOW5 \\
10:08:29.5 & 12:20:43 & 1099 & 16.33 & 15.62 & 15.31 & 1.02 & & ALW13 \\
10:08:30.0 & 12:16:04 & 1130 & 17.04 & 16.27 & 16.07 & 0.96 & m & MOW85 \\
10:08:31.0 & 12:17:08 & 1020 & 15.67 & 14.91 & 14.64 & 1.03 & & ALW14 \\
10:08:31.3 & 12:22:48 & 4011 & 16.39 & 15.68 & 15.45 & 0.95 & &  \\
10:08:31.5 & 12:18:54 & 1079 & 16.64 & 15.91 & 15.76 & 0.88 & m & MOW25 \\
10:08:32.5 & 12:18:06 & 1066 & 16.59 & 15.85 & 15.69 & 0.89 & m & MOW30 \\
10:08:33.6 & 12:15:33 & 1048 & 16.64 & 15.93 & 15.72 & 0.92 & &  \\
10:08:33.6 & 12:15:41 & 1049 & 16.39 & 15.63 & 15.43 & 0.97 & m & MOW113 \\
10:08:33.6 & 12:20:11 & 1094 & 16.53 & 15.75 & 15.57 & 0.96 & &  \\
10:08:34.4 & 12:15:56 & 1052 & 16.47 & 15.72 & 15.54 & 0.93 & m & MOW107 \\
10:08:34.9 & 12:19:39 & 1089 & 16.58 & 15.84 & 15.68 & 0.90 & m & MOW66 \\
10:08:34.9 & 12:21:03 & 1041 & 15.97 & 15.24 & 14.91 & 1.06 & m & MOW109,ALW16 \\
10:08:35.2 & 12:17:24 & 1021 & 16.22 & 15.48 & 15.23 & 0.99 & m & MOW68,ALW18 \\
10:08:36.9 & 12:20:11 & 1095 & 16.42 & 15.65 & 15.46 & 0.96 & m & MOW96 \\
10:08:37.1 & 12:19:24 & 1085 & 16.60 & 15.85 & 15.67 & 0.93 & m & MOW89 \\
10:08:38.0 & 12:15:52 & 1051 & 16.87 & 16.15 & 15.97 & 0.90 & m & MOW145 \\
10:08:38.3 & 12:15:45 & 1050 & 16.58 & 15.82 & 15.65 & 0.93 & m & MOW157 \\
10:08:39.0 & 12:18:32 & 1073 & 16.47 & 15.65 & 15.46 & 1.00 & &  \\
10:08:41.8 & 12:20:47 & 1040 & 16.07 & 15.29 & 15.15 & 0.92 & &  \\
10:08:42.1 & 12:19:32 & 1087 & 16.80 & 16.08 & 15.93 & 0.87 & m & MOW153 \\
10:08:42.8 & 12:23:45 & 3020 & 16.97 & 16.18 & 15.95 & 1.02 & m & MOW289 \\
10:08:44.2 & 12:23:46 & 3017 & 15.76 & 15.05 & 14.87 & 0.89 & &  \\
10:08:44.5 & 12:17:23 & 2027 & 16.74 & 15.97 & 15.79 & 0.95 & m & MOW179 \\
10:08:45.3 & 12:16:32 & 2016 & 16.03 & 15.26 & 15.09 & 0.94 & &  \\
10:08:46.0 & 12:22:33 & 3034 & 17.02 & 16.22 & 16.05 & 0.97 & m & MOW279 \\
10:08:48.3 & 12:19:42 & 2035 & 16.82 & 16.08 & 15.89 & 0.93 & m & MOW236 \\
10:08:55.5 & 12:14:01 & 9047 & 16.94 & 16.20 & 16.06 & 0.88 & m & MOW324 \\
\hline
\end{longtable}
\twocolumn
\end{center}

The chemical evolution of Leo~I has been discussed by many authors, most
recently by Gullieuszik et al. (2009), who use the calcium triplet to find a
very narrow distribution peaking at $\rm[M/H]= -1.2$. Very little is known
about the abundances of the $\alpha$-elements, but Shetrone et al. (2003)
suggest that the ratio of $\alpha$- to iron-peak elements is lower in
dwarf spheroidals, including Leo~I, than in globular clusters. If the
element enrichment pattern in Leo~I has been comparable to that in Fornax
(Letarte 2007) then $\alpha$-poor stars as old as 10\,Gyr may be present,
although those with ages comparable to the globular clusters will be 
$\alpha$-rich.

Estimates for the interstellar extinction towards Leo~I are low, and range
from $E(B-V)=0.01$ mag (Bellazzini et al. 2004) to $E(B-V)=0.04$ mag (Held et al.
2001).  Here we use a value of $E(B-V)=0.04$ mag, which amounts to
$A_K<0.01$ mag and therefore has a negligible effect on anything deduced from the
infrared photometry discussed here.

\begin{figure}
\includegraphics[width=8.5cm]{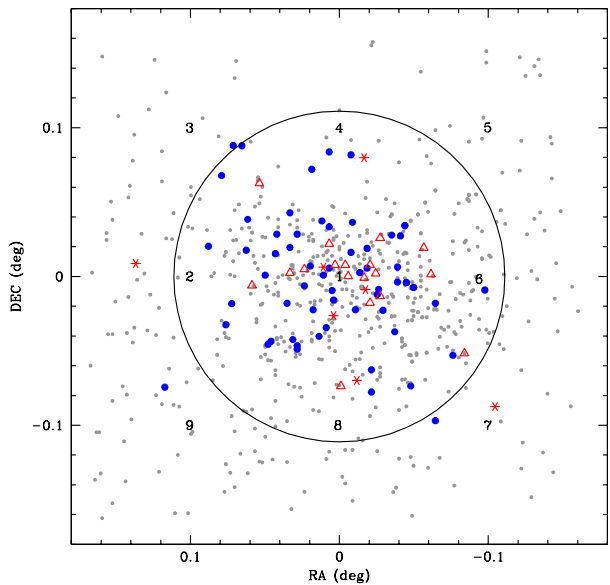}
\caption{Positions of stars measured in Leo~I. Field centres are shown by the numbers 1-9. Small grey dots
are red giant branch and field stars, large dots are probable AGB stars; Miras are shown as asterisks and other variables as triangles.
The circle represents a radius of 400 arcsec, within which the velocity distribution is isotropic (Mateo et al. 2008).}
\label{fig_pos}
\end{figure}

\section{Observations}
 Observations were made with the SIRIUS camera on the Japanese-South African
Infrared Survey Facility (IRSF) at SAAO, Sutherland. The camera produces
simultaneous $J$, $H$ and $K_S$ images covering an approximately $7.2 \times
7.2$ arcmin square field (after dithering) with a scale of 0.45
arcsec/pixel. Initially, only two adjacent fields, one centred on the galaxy
and one to the east, were observed (see Paper 1).  When it was realised that
there was a very red variable in the eastern field, the coverage was
extended to a $3\times 3$ grid. With generous overlaps between adjacent
fields, the area observed was 19.7$\times$19.2 arcmin squared (see
Fig.~\ref{fig_pos}).

Images were obtained at about 18 epochs spread over 3 years in fields 1
and 2, and between 10 and 15 epochs over about 2 years in the remaining
seven fields. In each field, 10 dithered images were combined after
flat-fielding and dark and sky subtraction. Typical exposures were of either
30 or 20\,s for each image, depending on the seeing and on the brightness of
the sky in the $K_S$ band. Photometry was performed using DoPHOT (Schechter
et al. 1993) in `fixed-position' mode, using the best-seeing $H$-band image
in each field as templates. Aladin was used to correct the coordinate system
on each template and RA and Dec were determined for each measured star. This
allowed a cross-correlation to be made with the 2MASS catalogue (Cutri et
al. 2003), and photometric zero points were determined by comparison of our
photometry with that of 2MASS. In each field, stars in common with the 2MASS
catalogue with photometric quality A in each colour were identified and the
IRSF zero point was adjusted to match that of 2MASS. The number of common
stars per field varies from 28 in the middle field to 6 in field 4 while
field 5 has only 2 stars, in common with
2MASS. The mean standard deviation over all fields of the differences
between IRSF and 2MASS are 0.04 mag in $J$ and 0.06 mag in each of $H$ and
$K_S$. No account was taken of possible colour transformations, such as in
Kato et al. (2007). Those transformations were derived using highly reddened
objects to define the red end and it is not obvious that the same
transformations will apply to carbon stars.

\section{Colour-Magnitude and Colour-Colour Diagrams}

\begin{figure} 
\includegraphics[width=8.5cm]{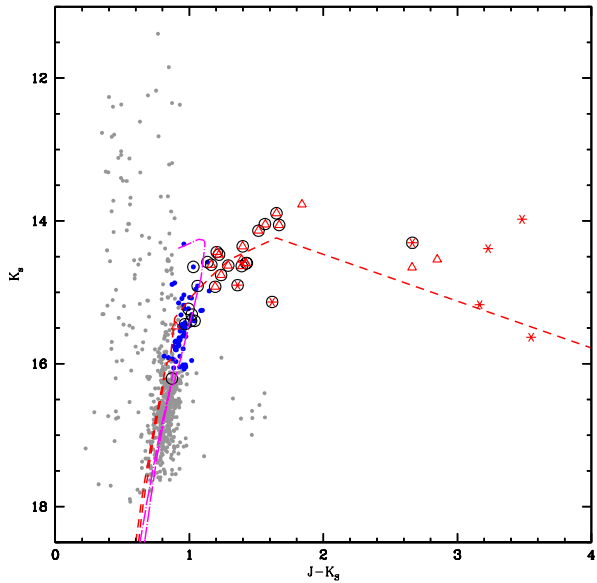}
\caption{Colour-magnitude diagram for Leo~I; Grey dots are field and red
giant branch stars, asterisks are Mira variables, triangles are low
amplitude, SR or Irr, variables, while large dots are other AGB stars.
Carbon stars are shown as symbols surrounded by
open circles. The dash-dotted and dashed curves are isochrones from Marigo
et al. (2008) for populations with ages of 2 and 10 Gyr, respectively
(details in text). } 
\label{fig_cm} 
\end{figure}

\begin{figure}
\includegraphics[width=8.5cm]{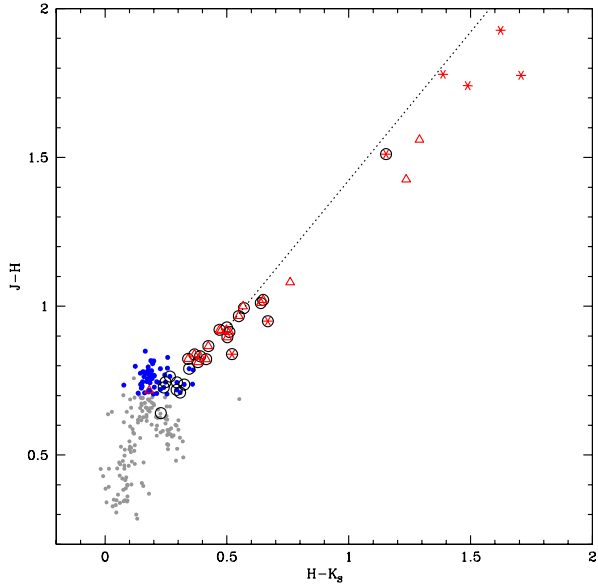}
\caption{Two colour diagram for Leo~I; symbols as in Fig.~\ref{fig_cm}. The dotted line represents the locus of
Galactic carbon Miras (see Whitelock et al 2009).}
\label{fig_jhhk}
\end{figure}

Fig.~\ref{fig_cm} shows the $K_S-(J-K_S)$ diagram and
Fig.~\ref{fig_jhhk} the $(J-H)-(H-K_S)$ diagram for all variables, which will
be discussed in section 4.2, and for constant stars 
selected on the basis of their standard deviations: 
for stars with $J<17.2$, $H<16$ or $K_S<16$ mag,
standard deviations $\sigma < 0.1$ mag; at the faint end 
the limit rises to 0.2 at $J=18.5$, and 0.25 at $H=K_S=18.0$.
Mean magnitudes from all of our observations are used in all of these plots.

The clump of points around $J-K_S\sim 1.5$ and $K_S\sim 16.6$ are
most likely unresolved background galaxies (see Whitelock et al. 2009).

To illustrate the possible evolutionary state of the stars plotted, we have
included two isochrones from Marigo et al. (2008) assuming a distance
modulus of 21.80 and a visual extinction of A$_V$=0.12. The first is for a
10 Gyr population with metallicity Z=0.001. This has a blue-ward hook at the
brightest magnitude which takes it into the region of the brightest blue
stars we have selected as AGB stars ($K_S<15, J-K_S\sim0.8$); the tip of its red
giant branch (TRGB) is at $K_S=15.85$ mag. The other isochrone is for an age
of 2 Gyr and Z=0.0012; its TRGB is at $16.20$ mag. This extends into the
region occupied by the red variables, but is very uncertain beyond
$J-K_S\sim1.6$.  The metallicity was chosen on the basis of the recent result
of Gullieuszik et al. (2009) showing a mean [M/H]=--1.2 (equivalent to
Z=0.0012) and very little dispersion. We have not included the isochrones on
the colour-colour diagram (Fig.~\ref{fig_jhhk}) since, as we found for
Fornax (Whitelock et al. 2008), there is very little correlation between the
isochrones and the data points.

\section{Asymptotic Giant Branch}
 We assume that stars with $K_S<16.1$ mag and $J-H>0.7$ are AGB stars in
Leo~I; $K_S=16.14$ mag is the TRGB (Held et al. 2010) and objects with
$J-H<0.68$ will be foreground dwarfs. In this way we make a conservative
selection and can be reasonably  certain that all stars selected will be AGB
stars in Leo~I, even if the selection also results in a few AGB stars being
omitted.  

From this selection we eliminated one star shown by Mateo et al. (2008) to
be a radial velocity non-member. The data for the AGB stars are listed in
Table 1, where the column headed $\rm L\#$ refers to our internal star numbers,
and the stars in common are noted by the Mateo et al. internal numbers
prefixed by MOW. Mateo et al. assigned stars to the AGB or RGB based on the
$VI$ colour-magnitude diagram. We find two of their AGB stars fall on our RGB,
both on the blue side of the RGB in our colour-magnitude diagram. A further
five stars that they call RGB stars are above our dividing line, four being
on the extreme red side of our AGB. There may be some physical reason for
these discrepancies rather than photometric error, but this is not really
relevant to the subject matter of the present paper.


\begin{table*}
\begin{center}
\caption{Periodic Red Variables.}
\label{tab_LPV}
\begin{tabular}{rcccccccccccl}
\hline
\multicolumn{1}{c}{L\#} & \multicolumn{1}{c}{P} & \multicolumn{1}{|c}{$J$} &  
\multicolumn{1}{c}{$\Delta J$} & \multicolumn{1}{c}{$H$} &
\multicolumn{1}{c}{$\Delta H$} &
\multicolumn{1}{c}{$K_S$}& \multicolumn{1}{c}{$\Delta K_S$} & $J-K_S$ &\multicolumn{1}{c|}{$m_{bol}$}& 
 Sp & 2MASS&other names\\
& \multicolumn{1}{c}{(day)}&\multicolumn{8}{c}{------------------------------------ (mag) -------------------------------------- }& & &  \\
\hline
{\bf Miras}\\

1019 & 158 &  16.75 & 1.21 & 15.80  & 0.89 &15.14 &   0.62 & 1.61 & 18.58 & C &10082751+1216539 & ALW9 \\
8026 & 180 &  16.30 & 0.78 & 15.40  & 0.61 &14.91 &   0.41 & 1.39 & 18.23 & C & 10082387+1214165 & ALW4\\
7020 & 191 &  18.34 & 1.87 & 16.56  & 1.57 &15.17 &   1.24 & 3.17 & 18.20 & &10080111+1213144\\
4005 & 252 &  16.97 & 0.55 & 15.46  & 0.53 &14.31 &   0.48 & 2.66 & 17.62 & C &10082268+1223159 & C13\\
2077 & 283 &  19.18 & 1.87 & 17.25  & 1.23 &15.63 &   1.17 & 3.53 & 18.33 & & & E\\ 
1077 & 336 &  17.62 & 1.23 & 15.88  & 1.01 &14.39 &   0.81 & 3.23 & 17.32 & &10082927+1218516& A\\
1064 & 523 &  17.46 & 1.52 & 15.68  & 1.29 &13.98 &   1.03 & 3.48 & 16.63 & &10082225+1217571& C\\
\multicolumn{10}{l}{\bf SRs} \\
6015 & 141 & 16.38 & 0.29 & 15.66 & 0.34 & 15.48 & 0.26 & 0.90 &   &  & 10080593+1215228 & MOW273 \\
1031 & 216 & 16.01 & 0.29 & 15.09 & 0.19 & 14.59 & 0.21 & 1.42 &  17.92 & C & 10082561+1218571 & C10 \\
1043 & 222 & 16.02 & 0.40 & 15.10 & 0.25 & 14.63 & 0.19 & 1.39 &  17.94 & C &10083988+1222144 & C08\\
1037 & 316 & 15.73 & 0.60 & 14.70 & 0.35 & 14.06 & 0.21 & 1.67 & 17.50 & C &10082008+1220023 & C02 \\
1067 & 999 & 17.31 & 0.43 & 15.89 & 0.31 & 14.65 & 0.29 & 2.66 & 17.92 && 10084120+1218068& D, MOW128\\
\hline
\end{tabular}
\end{center}
\flushleft{Notes: ABCDE sources identified in Paper I. ALW numbers from  Azzopardi, Lequeux \& Westerlund (1986). 
C numbers from  Demers \& Battinelli (2002). MOW numbers from Mateo et al. (2008).}
\end{table*}

\begin{table*}
\begin{center}
\caption{Variables without measured periods.}
\label{vars}
\begin{tabular}{rccccccccccl}
\hline
L\#& \multicolumn{1}{c}{$J$} &  ${\delta J}$ &
\multicolumn{1}{c}{$H$} & ${\delta H}$ &
\multicolumn{1}{c}{$K_S$}& ${\delta K_S}$& $J-K_S$& $m_{bol}$& Sp & 2MASS&other names\\
&\multicolumn{8}{c}{--------------------------------- (mag) --------------------------------- } & & & \\
\hline
1080  &    17.39 & 2.3 & 15.83 & 1.1 & 14.54 & 0.9& 2.85 & 17.73&  & 10082730+1218571& B\\
1012  &    15.54 & 0.3 & 14.53 & 0.3 & 13.89 & 0.3& 1.65 & 17.33&C & 10082064+1218364 & C03 \\
1013  &    15.65 & 0.4 & 14.69 & 0.2 & 14.14 & 0.3& 1.52 & 17.52&C & 10083233+1218460 & C07\\
1014  &    15.61 & 0.4 & 14.53 & 0.3 & 13.77 & 0.3& 1.84 & 17.25&  & 10082170+1218575 \\
1025  &    15.75 & 0.2 & 14.86 & 0.2 & 14.36 & 0.2& 1.40 & 17.68&C & 10082528+1218301& ALW8,C06 \\
1035  &    16.02 & 0.6 & 15.10 & 0.4 & 14.59 & 0.3& 1.42 & 17.93&C & 10081288+1219379 & C11 \\
6006  &    15.61 & 0.2 & 14.61 & 0.1 & 14.04 & 0.1& 1.56 & 17.44&C & 10081170+1218334  & C04\\
1022  &    15.91 & 0.3 & 15.05 & 0.2 & 14.62 & 0.2& 1.29 &&C & 10082175+1217249 & ALW3,C09\\ 
8024  &    16.12 & 0.3 & 15.30 & 0.2 & 14.92 & 0.2& 1.19 &&C & 10082634+1214026 & ALW6\\
6012  &    15.93 & 0.3 & 15.02 & 0.2 & 14.54 & 0.2& 1.39 & 17.85  && 10081288+1219379 &     \\
1036  &    16.00 & 0.3 & 15.17 & 0.2 & 14.76 & 0.2& 1.24 &&C & 10082849+1218485 & ALW11,C12\\
1011  &    15.69 & 0.2 & 14.86 & 0.1 & 14.47 & 0.1& 1.22 &&C & 10082254+1218259 & ALW5,C05\\
1023  &    15.64 & 0.3 & 14.80 & 0.2 & 14.44 & 0.2& 1.20 &&C & 10081995+1217414 & ALW2,C01\\
1028  &    15.78 & 0.3 & 14.96 & 0.2 & 14.61 & 0.2& 1.16 &&C & 10083471+1218374 & ALW15\\
\hline
\end{tabular}
\end{center}
\flushleft{Notes: other names follow the same convention as
Table~\ref{tab_LPV}.} 
\end{table*}

\subsection{Carbon stars}
 Azzopardi, Lequeux \& Westerlund (1986) and Demers \& Battinelli (2002)
identified 26 carbon stars in Leo~I and all of these appear in our sample; they
are identified in Figs.~\ref{fig_cm}, \ref{fig_jhhk} and in the tables. With
one exception the carbon stars are all brighter than $K_S=15.5$
($M_K=-6.3$). In the 2 Gyr isochrone, illustrated in Fig.~\ref{fig_cm}, stars
brighter than $K_S=15.7$ are carbon rich, while the 10 Gyr model does not
produce C-stars at all.

The faintest carbon star, ALW12 ($K_S=16.21$ mag), is referred to as
a `probable carbon star' by Azzopardi et al. (1986); its status therefore
remains uncertain. If it really is carbon rich it is too faint for this
enrichment to be the result of third dredge-up; it could, however, be
an extrinsic carbon star (i.e. one which received its enrichment by mass
transfer from a companion).
 
 Given the low metallicity and, more importantly, the probable low oxygen
abundance (Shetrone et al. 2003) of Leo~I we would anticipate that all of
the intermediate-age AGB variable stars will be carbon-rich. It is less
obvious what we should expect for very old AGB variables, but we do not
expect to see high mass-loss, i.e. very red, stars that are oxygen-rich in
this environment (Lagadec \& Zijlstra 2008).

The Miras L7020 and L2077 are outside the regions surveyed for carbon stars.
The other three periodic variables that have no carbon star identification
have $J-K_S>2.6$ (L6015 is different and is discussed in the next section);
presumably all these stars are extremely faint at short wavelengths.

Held et al. (2010) do a detailed division of AGB stars into presumed oxygen-
and carbon-rich on the basis of their colours and compare carbon star
numbers to theoretical predictions.

\subsection{Variable stars}
 We examined all of the bright red stars ($J-K_S>1.0$) with significant standard
deviations ($\sigma _{JHK_S}\geq0.1$ mag) for periodic variations. (Although it
is bluer than this limit, L6015, with
$J-K_S=0.9$, is relatively bright and was recognised as
a variable when its photometry was noticed to have much larger 
standard deviations in all wavebands than stars of similar brightness.)

Table \ref{tab_LPV} lists the Fourier mean $JHK_S$ magnitudes for the
periodic variables, the peak-to-peak amplitudes ($\Delta J,\ \Delta H,\
\Delta K_S$), mean $J-K_S$, and apparent bolometric magnitude, $m_{bol}$;
 a C indicates it
is a known carbon star, while 2MASS and other identifications are also given.  

\begin{figure}
\includegraphics[width=8.5cm]{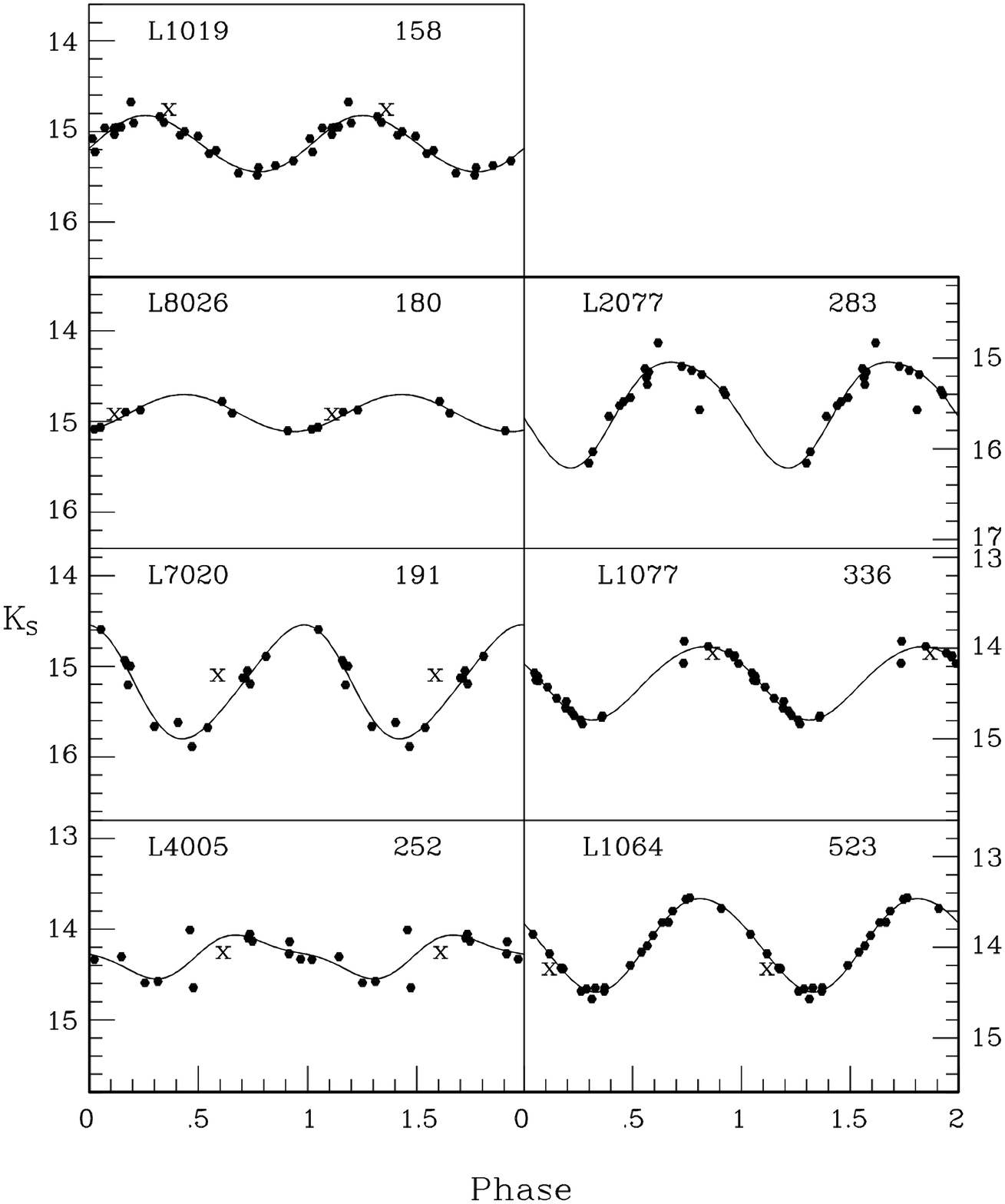}
\caption{The $K_S$ light curves for the Mira variables; each point is
plotted twice to emphasize the periodicity and the curves are the 
best-fitting second-order sine curves (first order for L8026). 
The X shows the 2MASS observation. The period in days is shown in the top right corner of each panel.}
\label{fig_lc}
\end{figure}

\begin{figure}
\includegraphics[width=8.5cm]{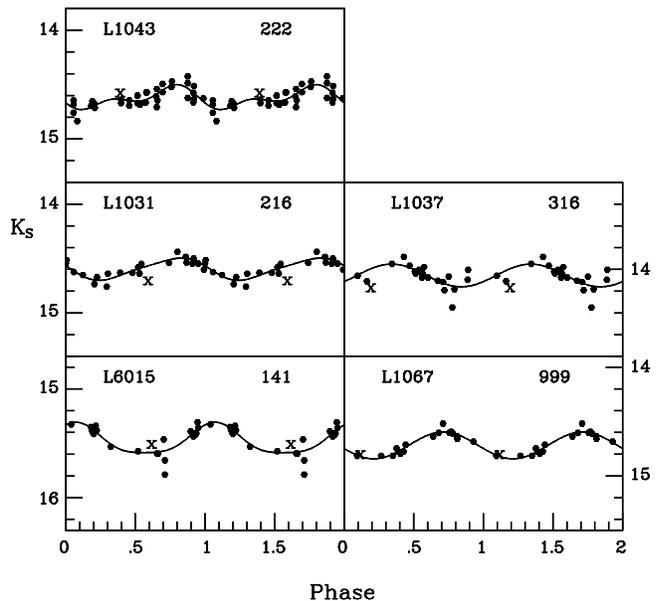}
\caption{The $K_S$ light curves for the SR variables; see Fig.~\ref{fig_lc}
for details. }
\label{fig_lcsr}
\end{figure}

Following Whitelock et al. (2006) we call variables with $\Delta K_S >0.4$ mag
Miras, and lower amplitude variables semi-regulars (SRs). Figs.~\ref{fig_lc}
and \ref{fig_lcsr} illustrate the light curves of the Mira and SR
variables, respectively, phased at their best-fitting periods.  Note that
the periods of the Miras are, in general, much more secure than those of the SR
variables.

With the exception of L4005 the Miras are all definite Miras. L4005 is included
with the group because of its large amplitude, but its light curve is not
well defined so its status as a Mira is uncertain. L8026 is almost
certainly a Mira variable, but its light curve is not well defined as only
8~$H$ observations (7~$K_S$) were available to define its period and mean
magnitude and they do not cover maximum light. L2077 is clearly a Mira, but
also has a long-term trend, it is discussed in Section 4.3.

Stars which are clearly variable, but for which we have been unable to
establish a period, are listed in Table~\ref{vars} together with the full
range of their variations: $\delta J,\ \delta H,\ \delta K_S$. Some of these
will be irregular variables, others are SRs, but require more observations
to define their periods. 

The variables are identified on the colour-magnitude
(Fig.~\ref{fig_cm}) and $(J-H)-(H-K_S)$ (Fig.~\ref{fig_jhhk}) diagrams. The
reddest stars (those with $J-K_S\geq 1.1$) are all variable. It is generally 
understood that AGB stars move to the right and down in Fig.~\ref{fig_cm} and to the upper right 
in Fig.~\ref{fig_jhhk} as their mass-loss
increases and their shells become optically thick at $JHK_S$.

The SR variable L6015, at $J-K_S\sim 0.9$ is significantly less red than any
of the other variables, which indicates that it has a very low mass-loss
rate. Nevertheless, the period established here puts it on the same PL($K$)
relation (see section 4.4) as the Miras and suggests that it is pulsating in
the fundamental mode (Wood 2000). This star is a radial velocity member of
Leo~I according to Mateo et al. (2008).

\begin{figure}
\includegraphics[width=8.5cm]{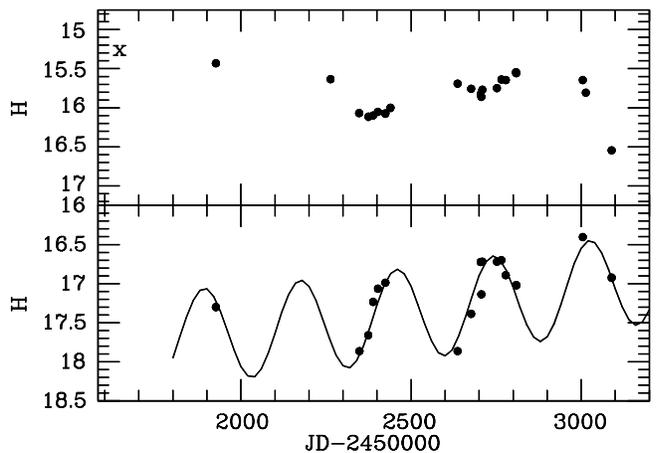}
\caption{The $H$ light curves for L1080 (top) and L2077 (bottom) are shown as a function of Julian Date 
(JD). For L2077 the fitted curve is a 280 day sine function plus a long term trend
represented as a 10000 day sine function; note that the mean light level
brightens by more than 0.5 mag over 5 cycles. L1080 shows large variations,
but no clear periodicity; X indicates the 2MASS measurement. }
\label{fig_2077}
\end{figure}

\subsection{Long term trends}
 Whitelock et al. (2006) established that among Galactic carbon-variables
about one third of Miras showed obscuration events, as did an unknown
fraction of non-Mira variables. An obscuration event is generally associated with the
ejection of a puff of dust into our line of sight, and the phenomenon is
particularly well illustrated by the light-curve of II Lup (Feast et al.
2003) which covers a time span of 18 years. The ejection of puffs of dust in
random directions is well studied among the hydrogen-deficient R CrB (RCB)
stars.

Fig.~\ref{fig_2077} illustrates two of the Leo~I variables that show obvious
long-term trends. L1080 is clearly variable, but not periodic. Its
light-curve is reminiscent of that of an RCB star, in that it shows SR-type
variability before going into a deep minimum. Its colours are very red for a
non-Mira and in the colour-magnitude diagram (Fig.~\ref{fig_cm}) it is found
among the extreme AGB stars with $J-K_S>2.5$.

L2077 is a Mira with a clear period of 280 days, but its mean magnitude also
brightened by over 0.5 mag at $H$ (by about 1.0 mag at $J$ and 0.5 mag at
$K_S$)
during the 3 years we monitored it. It is rather faint for its period as is
discussed in section 4.5.

Most of the observations by Held et al. (2010) for the variables are
completely consistent with the values in Tables~\ref{tab_LPV} and
\ref{vars}, the one exception being the SR variable L1037 (21484 in their
table 5), which we both identify with C02 from Demers \& Battinelli (2002).
They find $J=17.12,\ H=15.74\ K=14.83$, which is much redder than the values
in Table~\ref{tab_LPV} and the 2MASS observation and, as well, is outside of the
range illustrated in Fig. \ref{fig_lcsr}. The Held et al. observations were
made almost a year after ours and may indicate the ejection of a dust shell.

Whitelock et al. (2009) identified several variables in Fornax that seemed
to show examples of dust shell ejecta.

\subsection{Distribution of AGB stars and variables} 
 In Fig~\ref{fig_pos} we show where the AGB stars and variables lie with
respect to the centre of Leo~I. As found by Mateo et al. (2008), we see that
most of the AGB stars lie within a circle of radius 400 arcsec and are less
widely distributed than the RGB stars. Mateo et al. convincingly demonstrate
that the extended AGB cannot be from the same population as the bulk of the
GB stars. 

It is perhaps surprising that two of the Mira variables, L2077 and L7020,
lie well outside the 400 arcsec circle (they are 490.4 and 493.4 arcsec
from the centre, respectively). There are only three probable AGB stars
lying outside the circle, one of them being a confirmed radial velocity 
member.
It is possible that these two Miras, which have 
relatively short periods, and, by analogy,
the other three with short periods, are old stars and not from the same
population as the main extended AGB. This is discussed further in
section 6.

\subsection{Bolometric magnitudes and the period-luminosity (PL) relation}

 In calculating the bolometric magnitudes of the variables we assume
$E(B-V)=0.04$ mag, but note that the difference
in distance modulus derived from this assumption and $E(B-V)=0.00$ mag
is less than 0.01 mag.

\begin{figure}
\includegraphics[width=8.5cm]{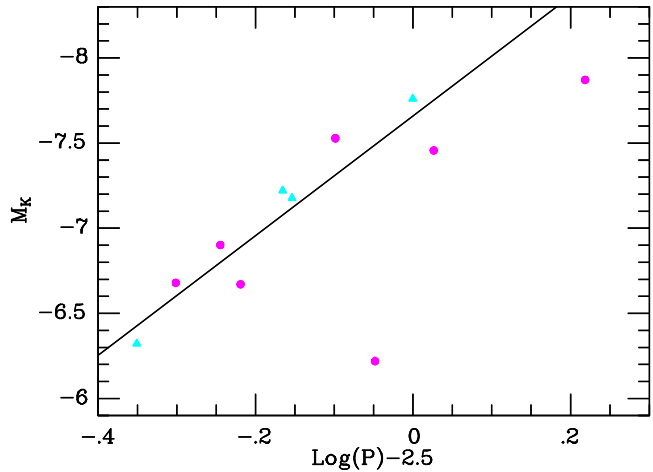}
\caption{The $M_K$ PL relation for the periodic variables in Leo I
on the assumption that the distance modulus is 21.80 mag;
Miras are shown as circles and SRs as triangles. The line is the $K$ PL
relation derived from C-rich LMC Miras.}
\label{fig_kpl}
\end{figure}

Fig. \ref{fig_kpl} shows the PL($K$) relation for the periodic variables in
Leo~I compared to the relation for C-rich Miras in the LMC: 
\begin{equation} M_K=-3.51[\log P-2.38]-7.24, \end{equation} 
from Whitelock, Feast \& van Leeuwen (2008) on the assumption
that the distance modulus of Leo~I is 21.80 (see below). Five of the Miras
fall close to the relation and the other two considerably below it. As these
stars are red compared to the LMC objects used to define the PL relation,
this is hardly surprising and we would anticipate that they are affected by
circumstellar extinction at $K$.

\begin{figure}
\includegraphics[width=8.5cm]{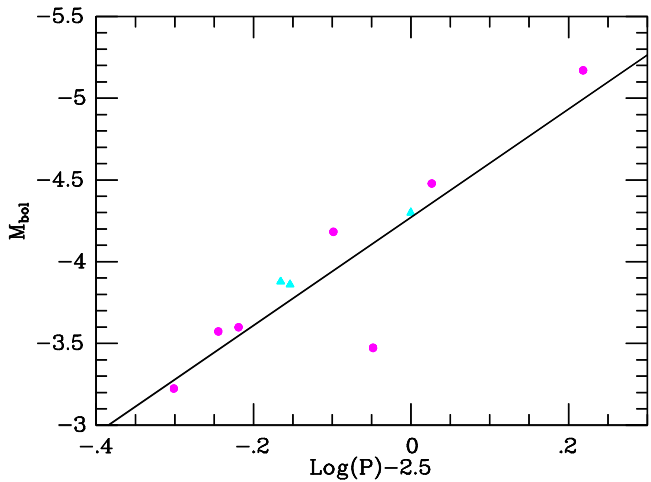}
\caption{The bolometric PL relation for the periodic variables in Leo I.
Miras are shown as circles and SRs as triangles. The line is the result of
fitting equation A1 from Whitelock et al. (2009) to all 6 
Miras, which gives a distance modulus of 21.80 mag.}
\label{fig_bol}
\end{figure}

Apparent bolometric magnitudes were calculated from colour dependent bolometric
corrections, as was done in our previous work on Fornax (Whitelock et al. 2009). 
To determine the distance to Leo~I we fit the following equation
to the Mira observations:
\begin{equation}M_{bol}=-4.271-3.31[\log P -2.5] \end{equation} (This is A1
from Whitelock et al. (2009), and assumes an LMC modulus of 18.39.)  Using all 7 Miras we
obtain $(m-M)=21.80\pm0.11$ mag (internal error only). Eliminating L4005
(which has an uncertain status as a Mira) and using the other 6 stars gives
$(m-M)=21.84\pm0.12$ mag. Eliminating L2077 (which was undergoing an
obscuration event when we observed it, see section 4.3) and using the other
6 stars gives $(m-M)=21.69\pm0.04$ mag. Fig.~\ref{fig_bol} shows the PL
relation with all the variables and the relation just derived for all 7
Miras.

Note that the SR variable L6015 does not appear in Fig.~\ref{fig_bol};
we cannot use the same method to estimate its bolometric magnitude
because the bolometric corrections are not defined for stars with
$J-K<1.4$ (on the SAAO system).

Table~\ref{vars} includes estimates of the bolometric magnitudes for the
variables without periods where their colours allow it to be estimated, but
these should be used with caution as we do not know that these variables are
comparable to those for which the bolometric correction was established.

To establish the complete (external plus internal) error on the distance
modulus we must include the uncertainty of the adopted distance modulus for
the LMC, $18.39 \pm 0.05$ (van Leeuwen et al.2007, derived from Cepheid
variables). Taking this into account the distance modulus of Leo I from all
seven Miras is, $(m-M)_0 = 21.80 \pm 0.12$.

Omitting  L2077, on the assumption that its apparent low luminosity results
from obscuration by a non-uniform shell, results in a modulus of $(m-M)_0 =
21.69 \pm 0.06$. However, in view of the selection process involved, we have
increased the error in this case by adding in quadrature the, primarily
intrinsic, rms scatter per star. Whitelock et al. (2008) found the scatter
for LMC Miras about the PL to be $\pm 0.12$ mag, so we estimate the rms
scatter per star for the 6 stars here as $0.12/\sqrt{6}=0.049$. This
increases the standard error from 0.06 to 0.08.


\section{The Distance to Leo I}
 Here we compare the distance obtained above with independent estimates from
the literature using RR Lyrae variables (RRs) and the tip of the red giant
branch. These are expressed in a self consistent way, assuming the
interstellar reddening amounts to $E(B-V)=0.04$ mag (which affects estimates
based on shorter wavelength measurements although its impact on our infrared
measures is minimal). The results are given in Table~\ref{distance}.

Held et al. (2001) discuss $B$ and $V$ photometry for RRs they 
discovered in Leo~I. Assuming $E(B-V)=0.04$ mag for the galaxy, they find a
mean value for the RRs of $V_0=22.48 \pm 0.12$. We assume the following
expression for the absolute magnitude of the RRs (Feast
2010):\begin{equation} M_V=0.21(\rm [Fe/H]+1.5)+0.68 \ (\pm 0.10),
\end{equation} which was derived from a mean of the trigonometrical,
pulsation and statistical parallaxes. 
Held et al. adopted $\rm[Fe/H]=-1.82$,
giving $M_V=+0.61$. Hence $(m-M)_0=22.48-0.61=21.87\pm 0.14$ (standard error
from Held et al.). Equation 3, with the mean [Fe/H] value for the LMC from
Gratton et al. (2004), gives a distance modulus for the LMC of 18.38 mag.
Thus, on a scale of $(m-M)_0=18.39$ for the LMC, the RRs give $(m-M)_0=21.88
\pm 0.14$ for Leo~I. 

However, Gullieuszik et al. (2009) recently determined the metallicity of
Leo~I as $\rm [M/H]=-1.2$, with a very narrow intrinsic dispersion of only
$\pm0.08$. Converting this to [Fe/H] depends on the metallicity scale and is
rather uncertain, but using the globular cluster comparison made by
Gullieuszik et al., we would estimate this corresponds to $\rm [Fe/H]\sim -1.4$.
Following the same argument as above this leads to $M_V=0.70$ for the RRs
and $(m-M)_0=21.79 \pm 0.14$ mag for the distance modulus.

\begin{table}
\begin{center}
\caption{Distance to Leo I.}
\label{distance}
\begin{tabular}{lcl}
\hline
method & $(m-M)_0$ & reference\\
 & (mag) &\\
\hline
RRs & $21.88\pm 0.14$ & Held et al. (2001) revised\\
     & $21.79\pm0.14$ & see text\\
TRGB($I$) & $21.97\pm 0.13 $ & Bellazzini (2004) revised\\
TRGB($JHK$) & $21.88\pm0.13$ & Held et al. (2010) revised\\
Miras & $21.80\pm 0.12$ & this paper \\
\hline
\end{tabular}
\end{center}
\end{table}

Bellazzini et al. (2004) discuss the distance to Leo~I from the TRGB, using
an independent calibration of $M_I^{TRGB}$ based on adopted distances of
$\omega$~Cen and 47~Tuc. They find $I{\rm(TRGB)}=17.97$.  Assuming
$E(B-V)=0.01$ mag and $\rm [M/H]\sim -1.2$ they obtain
$(m-M)_0=22.02\pm0.13$. With our adopted value of $E(B-V)=0.04$ mag, this
becomes $(m-M)_0=21.97\pm0.13$.  Held et al. (2010) derive a modulus of
$22.04 \pm 0.11$ from the TRGB at $JHK$ after applying population corrections.
As in the case of Fornax discussed by Whitelock et al. (2009) their basic
scale is 0.16 mag longer than that adopted here. Thus on a scale consistent
with the other results in Table 4 the infrared TRGB distance is 21.88. Note,
however, Salaris \& Girardi (2005) have urged caution in using the TRGB as a
distance indicator in the presence of a significant intermediate mass
population --- a situation that obviously applies to Leo~I.

Each method of distance determination has its own uncertainties, and
although touched upon here the details of these are beyond the scope of this
paper. Within those uncertainties the various distance estimates to Leo~I agree
remarkably well and the Miras provide a useful estimate of the
distance, independent of the more commonly used methods. If the metallicity
of the C-Miras in Leo~I is lower than that in the LMC (which seems likely) this
result indicates that there are no large metallicity effects in the Mira PL relation.

\section{Discussion}

  Our most direct information about the ages and masses of carbon 
Miras comes from those discovered in Magellanic Cloud clusters by Nishida et al.
(2000). These have periods of between 450 and 526 days and follow the same
bolometric PL relation as the carbon-rich field Miras (Whitelock et 
al. 2003). The clusters in which they are found have ages of about 1.6 Gyr
(Mucciarelli et al. 2007a; Mucciarelli, Origlia \& Ferraro 2007b; Glatt et
al. 2008) and we would therefore suggest that the longest period Mira in
Leo~I, L1064 with P=523 days, is of similar age.

The other Miras must be older; how much older is difficult to quantify. In
view of the fact that two of the stars with $P<300$ days are found in the
outer part of Leo~I (section 4.4) it is tempting to suggest that they may be
much older, possibly comparable to the 10 Gyr or more that is thought to be
characteristic of the RGB stars. Note that in the Galaxy oxygen-rich Miras
with $P<300$ days are found in relatively metal-rich ($\rm [Fe/H]>-1$)
globular clusters, which have ages greater than 10\,Gyr. Within these
clusters, their periods and therefore of course their magnitudes are
proportional to the metallicity of the parent cluster (Feast, Whitelock \&
Menzies 2002). Miras are not found in the more metal-deficient clusters.

Although we have yet to confirm spectroscopically that these two Miras are
carbon-rich, their colours certainly suggest it. Most models do not 
produce carbon stars or high-mass-loss objects at ages of 10 Gyr, although recent
work suggests it might happen. Karakas (2010) modelled a $\rm 1 M_{\odot}$ 
star with Z=0.0001 and found that it experienced 26 thermal pulses and a small
amount of third-dredge-up. In an envelope with such a low metallicity, 
even a small amount of dredge-up was enough to make $\rm C/O> 1$ and 
produce a carbon star. This is clearly an area where more work is needed and these
stars are worth a more detailed investigation.

For the future, with the next generation of large telescopes working in the
infrared, Mira variables will prove vital distance indicators for studying
populations of old and intermediate age stars, where they will be amongst
the most luminous objects, easily identified via their large amplitude
variations.

\section*{Acknowledgments} We are grateful to the following colleagues for
acquiring images for this programme: Toshihiko Tanab\'e,
Takahiro Naoi, Shogo Nishiyama, Yoshifusa Ita and Barbara Cunow.
 
This research has made use of Aladin. 
This publication makes use of data products from the Two Micron All Sky
Survey, which is a joint project of the University of Massachusetts and the
Infrared Processing and Analysis Center/California Institute of Technology,
funded by the National Aeronautics and Space Administration and the National
Science Foundation.
This material is based upon work supported financially by the South African National Research Foundation.
We also thank Enrico Held for sending us his 2010 paper in advance of
publication.

\end{document}